\documentclass[amssymb,twocolumn,eqsecnum,aps]{revtex4}
\usepackage{amsmath}
\usepackage{enumerate}
\usepackage{graphicx}
\begin{document}
\def \bbeta {{\beta}\hskip -6.6pt {\beta}}
\def \brho {{\rho} \hskip -4.5pt { \rho}}
\def \bnab {{\bf\nabla}\hskip-8.8pt{\bf\nabla}\hskip-9.1pt{\bf\nabla}}
\title{Alternative formulation of the macroscopic field equations in a
linear magneto-dielectric medium: Field equations}
\author{Michael E. Crenshaw}
%\email{michael.e.crenshaw4.civ@mail.mil, Phone: 1-256-876-3526}
\affiliation{Charles M. Bowden Research Laboratory, US Army Combat Capabilities Development Command (DEVCOM) - Aviation and Missle Center,
Redstone Arsenal, AL 35898, USA}
\begin{abstract}
We derive an alternative formulation of the field equations for
macroscopic electromagnetic fields in a linear magneto-dielectric medium
as an identity of the Maxwell--Minkowski equations, complementing a
variety of other representations including the Amp\`ere, Chu, Lorentz,
and Minkowski formulations of continuum electrodynamics.
In the new formulation of the macroscopic field equations, the material
properties are carried as a renormalization of the temporal and spatial
coordinates instead of as independent material constants.
The new representation of the field equations raises some interesting
physical issues with relativity and boundary conditions.
%\vskip 0.25cm
%keywords:
%continuum electrodynamics, macroscopic Maxwell equations
%\hfill
%\vskip 0.25cm
%Conflict of Interest: The author declares that there is no conflict of
%interest.
\end{abstract}
\maketitle
\par
\section{Introduction}
\par
The equations of motion for macroscopic electromagnetic fields
in a transparent linear magneto-dielectric medium are not unique.
In \textit{A Dynamical Theory of the Electromagnetic Field,} Maxwell
cast his theory of electromagnetism in terms of 20 quantities and 20
equations \cite{BISimpson}.
A decade later, Maxwell had reformulated his theory and the 28
``fundamental equations ... are scattered throughout more than twenty
pages of text spanning two chapters'' \cite{BIWheeler} of \textit{A
Treatise on Electricity and Magnetism.}
The vector description of electrodynamics was later introduced by 
Heaviside.
In a polarizable, magnetizable linear medium, the familiar vector field
equations \cite{BIJackson,BIMar,BIGriffiths,BIZangwill},
\begin{subequations}
\begin{equation}
\nabla\times{\bf H}-\frac{1}{c}\frac{\partial{\bf D}}{\partial t}=0
\label{EQt1.01a}
\end{equation}
\begin{equation}
\nabla\cdot{\bf B}=0
\label{EQt1.01b}
\end{equation}
\begin{equation}
\nabla\times{\bf E}+\frac{1}{c}\frac{\partial{\bf B}}{\partial t}=0
\label{EQt1.01c}
\end{equation}
\begin{equation}
\nabla\cdot{\bf D}=0 \, ,
\label{EQt1.01d}
\end{equation}
\label{EQt1.01}
\end{subequations}
are known as the Maxwell--Minkowski equations, the Maxwell--Heaviside 
equations, or the macroscopic Maxwell equations, although this is often
shortened to simply `the Maxwell equations.'
Here, the macroscopic fields, ${\bf E}$, ${\bf D}$, ${\bf B}$,
and ${\bf H}$, are functions of position, ${\bf r}$, and time, $t$.
The transparent linear medium is treated as being source free in order
to focus on the fundamental physical issues.
\par
There are other formulations of the macroscopic Maxwell equations that
are used to emphasize various features of continuum electrodynamics.
In the Chu formalism of
electrodynamics \cite{BIPenHau,BIKemp,BIKemplatest}, for example,
\begin{subequations}
\begin{equation}
\nabla\times{\bf H}^c-\frac{1}{c}\frac{\partial{\bf E}^c}{\partial t}=
\frac{1}{c}\frac{\partial{\bf P}^c}{\partial t}
\label{EQt1.02a}
\end{equation}
\begin{equation}
\nabla\cdot{\bf H}^c=-\nabla\cdot{\bf M}^c
\label{EQt1.02b}
\end{equation}
\begin{equation}
\nabla\times{\bf E}^c+\frac{1}{c}\frac{\partial{\bf H}^c}{\partial t}=
-\frac{1}{c}\frac{\partial{\bf M}^c}{\partial t}
\label{EQt1.02c}
\end{equation}
\begin{equation}
\nabla\cdot{\bf E}^c= -\nabla\cdot{\bf P}^c \, ,
\label{EQt1.02d}
\end{equation}
\label{EQt1.02}
\end{subequations}
the material response is separated from the electric field ${\bf E}^c$
and the magnetic field ${\bf H}^c$.
The Amp\`ere and Lorentz formulations of the field equations are also
extant in the recent scientific
literature \cite{BIPenHau,BIKemp,BIKemplatest}.
\par
In the current work, we derive another formulation of the Maxwell
equations of motion of macroscopic fields in a transparent linear
magneto-dielectric medium as a carefully derived identity of the
Maxwell--Minkowski equations.
Mathematical identities are ordinarily mundane, but it turns out that the
new representation of the field equations raises some interesting physical
issues with relativity and boundary conditions.
\par
\section{Maxwellian Continuum Electrodynamics}
\par
We define a simple linear medium as an idealized model of a medium that
can be treated as being at rest in the local coordinate frame and as
having a real, time-independent linear permittivity $\varepsilon({\bf r})$
and a real, time-independent linear permeability $\mu({\bf r})$
corresponding to the frequency of a monochromatic field, or to the center
frequency of a quasimonochromatic field \cite{BIAxiom}.
As the field enters the medium from the vacuum, the field imparts optically
induced forces to the material.
However, it takes an intense light field applied for a long time for the
material to be accelerated to relativistic speeds.
Then the non-relativistic constitutive relations
\begin{subequations}
\begin{equation}
{\bf D}=\varepsilon ({\bf r}){\bf E}
\label{EQt2.01a}
\end{equation}
\begin{equation}
{\bf B}=\mu ({\bf r}){\bf H}
\label{EQt2.01b}
\end{equation}
\label{EQt2.01}
\end{subequations}
are a valid limiting case and very good approximation.
(Relativistic corrections are of the order of $|{\bf v}|/c$.)
The constitutive relations, Eqs.~(\ref{EQt2.01}), are explicit axioms of
the theory.
\par
We substitute the usual constitutive relations, Eqs.~(\ref{EQt2.01}),
into the Maxwell--Minkowski equations, Eqs.~(\ref{EQt1.01}), to obtain
\begin{subequations}
\begin{equation}
\nabla\times{\bf H}
-\frac{1}{c}\frac{\partial (\varepsilon {\bf E})}{\partial t}=0
\label{EQt2.02a}
\end{equation}
\begin{equation}
\nabla\cdot (\mu{\bf H})=0
\label{EQt2.02b}
\end{equation}
\begin{equation}
\nabla\times{\bf E}+\frac{1}{c}\frac{\partial(\mu{\bf H})}{\partial t}=0
\label{EQt2.02c}
\end{equation}
\begin{equation}
\nabla\cdot(\varepsilon {\bf E})=0 
\label{EQt2.02d}
\end{equation}
\label{EQt2.02}
\end{subequations}
for the dynamics of macroscopic fields in a simple linear medium.
Let
\begin{subequations}
\begin{equation}
n_e({\bf r})=\sqrt{\varepsilon}
\label{EQt2.03a}
\end{equation}
\begin{equation}
n_m({\bf r})=\sqrt{\mu}
\label{EQt2.03b}
\end{equation}
\label{EQt2.03}
\end{subequations}
and substitute the time-independent
parameters $n_e({\bf r})$ and $n_m({\bf r})$
into Eqs.~(\ref{EQt2.02}).
We algebraically obtain
\begin{subequations}
\begin{equation}
n_m\nabla\times {\bf H}
-\frac{n_m}{c}\frac{\partial (n_en_e{\bf E})}{\partial t}= 0
\label{EQt2.04a}
\end{equation}
\begin{equation}
\nabla\cdot \left ( n_mn_m {\bf H} \right ) =0
\label{EQt2.04b}
\end{equation}
\begin{equation}
n_e\nabla\times{\bf E}
+\frac{n_e}{c}\frac{\partial (n_mn_m{\bf H})}{\partial t}=0
\label{EQt2.04c}
\end{equation}
\begin{equation}
\nabla\cdot (n_en_e{\bf E})=0 \, .
\label{EQt2.04d}
\end{equation}
\label{EQt2.04}
\end{subequations}
Next we use common vector
identities \cite{BIJackson,BIMar,BIGriffiths,BIZangwill} to commute the
material parameters, $n_e({\bf r})$ and $n_m({\bf r})$, with the vector
differential operators obtaining
\begin{subequations}
\begin{equation}
\frac{\nabla}{n_m}\times \left ( n_m{\bf H}\right )
+\frac{n_e}{c}\frac{\partial (-n_e{\bf E})}{\partial t}
= \frac{\nabla n_m}{n_mn_m}\times \left ( n_m{\bf H} \right )
\label{EQt2.05a}
\end{equation}
\begin{equation}
\frac{\nabla}{n_m}\cdot \left ( n_m{\bf H} \right )
= -\frac{\nabla n_m}{n_mn_m}\cdot \left ( n_m{\bf H} \right )
\label{EQt2.05b}
\end{equation}
\begin{equation}
\frac{\nabla}{n_m}\times (-n_e{\bf E})
-\frac{n_e}{c}\frac{\partial (n_m{\bf H})}{\partial t}
= \frac{\nabla n_e}{n_mn_e}\times (-n_e {\bf E})
\label{EQt2.05c}
\end{equation}
\begin{equation}
\frac{\nabla}{n_m}\cdot (-n_e{\bf E})
= -\frac{\nabla n_e}{n_m n_e}\cdot (-n_e {\bf E}) \, .
\label{EQt2.05d}
\end{equation}
\label{EQt2.05}
\end{subequations}
We reparamaterize the macroscopic electric and magnetic fields 
\begin{subequations}
\begin{equation}
{\bf \Pi}=-n_e({\bf r}){\bf E}
\label{EQt2.06a}
\end{equation}
\begin{equation}
{\bf \bbeta}=n_m({\bf r}){\bf H} 
\label{EQt2.06b}
\end{equation}
\label{EQt2.06}
\end{subequations}
and substitute the relations, Eqs.~(\ref{EQt2.06}), into
Eqs.~(\ref{EQt2.05}) such that
\begin{subequations}
\begin{equation}
\frac{\nabla}{n_m}\times \bbeta
+\frac{n_e}{c}\frac{\partial {\bf \Pi}}{\partial t}
= \frac{\nabla n_m}{n_mn_m}\times \bbeta
\label{EQt2.07a}
\end{equation}
\begin{equation}
\frac{\nabla}{n_m}\cdot \bbeta = -\frac{\nabla n_m}{n_mn_m}\cdot\bbeta
\label{EQt2.07b}
\end{equation}
\begin{equation}
\frac{\nabla}{n_m} \times{\bf \Pi}
-\frac{n_e}{c}\frac{\partial\bbeta}{\partial t}
= \frac{\nabla n_e}{n_mn_e}\times {\bf \Pi}
\label{EQt2.07c}
\end{equation}
\begin{equation}
\frac{\nabla}{n_m} \cdot {\bf \Pi}
= -\frac{\nabla n_e}{n_mn_e}\cdot {\bf \Pi} \, .
\label{EQt2.07d}
\end{equation}
\label{EQt2.07}
\end{subequations}
Equations~(\ref{EQt2.07}) are identities of the
macroscopic Maxwell equations, Eqs.~(\ref{EQt2.02}), for a linear 
magneto-dielectric medium.
As long as we use the common and well-established model of a 
monochromatic/quasimonochromatic field propagating through a simple
linear medium, Eqs.~(\ref{EQt2.07}) constitute a valid alternative
formulation of the macroscopic Maxwell field equations with the same
regime of applicability including the same material parameters, the same
boundary conditions, and the same limiting cases.
It is straightforward, although cumbersome, to apply this procedure to
more complex models of the medium.
\par
We write a new timelike coordinate
\begin{equation}
\bar x^0= \frac{x^0}{n_e}=\frac{ct}{n_e}
\label{EQt2.08}
\end{equation}
and note that $n_e$ is not explicitly a function of time for the simplified
model of a linear medium that we are treating here.
We substitute Eq.~(\ref{EQt2.08}) into
Eqs.~(\ref{EQt2.07}) to produce
\begin{subequations}
\begin{equation}
\frac{\nabla}{n_m}\times \bbeta
+\frac{\partial {\bf \Pi}}{\partial \bar x^0}
= \frac{1}{n_m}\frac{\nabla n_m}{n_m}\times \bbeta
\label{EQt2.09a}
\end{equation}
\begin{equation}
\frac{\nabla}{n_m}\cdot \bbeta
= -\frac{1}{n_m}\frac{\nabla n_m}{n_m}\cdot\bbeta
\label{EQt2.09b}
\end{equation}
\begin{equation}
\frac{\nabla}{n_m} \times{\bf \Pi}
-\frac{\partial\bbeta}{\partial\bar x^0}
= \frac{1}{n_m}\frac{\nabla n_e}{n_e}\times {\bf \Pi}
\label{EQt2.09c}
\end{equation}
\begin{equation}
\frac{\nabla}{n_m} \cdot {\bf \Pi}
= -\frac{1}{n_m}\frac{\nabla n_e}{n_e}\cdot {\bf \Pi} \, .
\label{EQt2.09d}
\end{equation}
\label{EQt2.09}
\end{subequations}
We introduce the notation
\begin{equation}
\bar \nabla=\left ( \frac{1}{n_m}\frac{\partial}{\partial x},
\frac{1}{n_m} \frac{\partial}{\partial y},
 \frac{1}{n_m}\frac{\partial}{\partial z} \right ) 
\label{EQt2.10}
\end{equation}
such that
\begin{subequations}
\begin{equation}
\bar\nabla\times \bbeta
+\frac{\partial {\bf \Pi}}{\partial \bar x^0}
= \frac{\bar\nabla n_m}{n_m}\times \bbeta
\label{EQt2.11a}
\end{equation}
\begin{equation}
\bar\nabla\cdot \bbeta = -\frac{\bar\nabla n_m}{n_m}\cdot\bbeta
\label{EQt2.11b}
\end{equation}
\begin{equation}
\bar\nabla \times{\bf \Pi}
-\frac{\partial\bbeta}{\partial\bar x^0}
= \frac{\bar\nabla n_e}{n_e}\times {\bf \Pi}
\label{EQt2.11c}
\end{equation}
\begin{equation}
\bar\nabla \cdot {\bf \Pi}
= -\frac{\bar\nabla n_e}{n_e}\cdot {\bf \Pi} \, .
\label{EQt2.11d}
\end{equation}
\label{EQt2.11}
\end{subequations}
Like Eqs.~(\ref{EQt2.09}), the set of Eqs.~(\ref{EQt2.11}) is a
systematically and rigorously derived identity of the
macroscopic Maxwell equations, Eqs.~(\ref{EQt2.02}), and constitutes a
valid alternative formulation of the Maxwell field equations for a
simple linear magneto-dielectric medium.
Each linear, isotropic, homogeneous medium will have a different set
of macroscopic field equations.
\par
Although we can retain the spatial dependence of $n_m$ and $n_e$, we 
will adopt the limit of an arbitrarily large, transparent, linear,
isotropic, homogeneous, magneto-dielectric medium in order to proceed
with the fundamental physical issues without unnecessarily complicated
formulas.
Although a real-world material cannot be of infinite extent, an
arbitrarily large medium can be treated as infinite in the sense that
light cannot reach the boundaries of the medium in the finite time that
it takes to perform an experiment.
For this limit, we introduce new spatial variables
\begin{subequations}
\begin{equation}
\bar x=n_m x
\label{EQt2.12a}
\end{equation}
\begin{equation}
\bar y=n_m y
\label{EQt2.12b}
\end{equation}
\begin{equation}
\bar z=n_m z 
\label{EQt2.12c}
\end{equation}
\label{EQt2.12}
\end{subequations}
such that Eqs.~(\ref{EQt2.11}) become
\begin{subequations}
\begin{equation}
\bar\nabla\times \bbeta +\frac{\partial{\bf \Pi}}{\partial\bar x^0} = 0
\label{EQt2.13a}
\end{equation}
\begin{equation}
\bar\nabla\cdot \bbeta = 0
\label{EQt2.13b}
\end{equation}
\begin{equation}
\bar\nabla\times{\bf \Pi} -\frac{\partial\bbeta}{\partial \bar x^0}=0
\label{EQt2.13c}
\end{equation}
\begin{equation}
\bar\nabla\cdot {\bf \Pi} = 0 \, ,
\label{EQt2.13d}
\end{equation}
\label{EQt2.13}
\end{subequations}
where the material Laplacian operator, Eq.~(\ref{EQt2.10}),
simplifies to
\begin{equation}
\bar \nabla=\left ( \frac{\partial}{\partial \bar x},
\frac{\partial}{\partial \bar y},
\frac{\partial}{\partial \bar z} \right ) 
\label{EQt2.14}
\end{equation}
for a linear, isotropic, homogeneous medium.
Each linear, isotropic, homogeneous medium will have a different set of
macroscopic field equations for each set of magnetic and dielectric
material characteristics.
\par
The wave equation
\begin{equation}
\bar \nabla\times(\bar \nabla\times{\bf A})
+\frac{\partial}{\partial \bar x^0}
\left ( \frac{\partial {\bf A}}{\partial \bar x^0}\right ) =0
\label{EQt2.15}
\end{equation}
is easily formed by substituting 
\begin{equation}
\beta=\bar\nabla\times{\bf A}
\label{EQt2.16}
\end{equation}
\begin{equation}
\Pi=\frac{\partial {\bf A}}{\partial \bar x^0}=0
\label{EQt2.17}
\end{equation}
into the variant Maxwell--Amp\`ere Law, Eq.(\ref{EQt2.13a}).
There is a different wave equation for each simple linear material, just
like there is in the Maxwell--Minkowski formulation
\begin{equation}
\nabla\times(\nabla\times{\bf A})
+\frac{n^2}{c^2}\frac{\partial^2 {\bf A}}{\partial t^2}=0  \, .
\label{EQt2.18}
\end{equation}
\par
\section{Significance}
\par
There is an extensive body of work in classical continuum electrodynamics
and in the relationship between continuum electrodynamics and other
physical principles.
Equations~(\ref{EQt2.11}) are obviously inconsistent with extant treatments
of Fresnel boundary conditions, special relativity, conservation, Lorentz
invariance, and other well-known physical principles in a linear medium.
Then, it is straighforward to use physical arguments from the prior work to
``prove'' that Eqs.~(\ref{EQt2.11}) are false.
However, such a proof would be a post hoc ergo propter hoc logical fallacy.
Significantly, Eqs.~(\ref{EQt2.11}) and (\ref{EQt2.02}) are identities and
disproving Eqs.~(\ref{EQt2.11}) by physical arguments would disprove the
Maxwell--Minkowski field equations in a dielectric, Eqs.~(\ref{EQt2.02}),
thereby negating the basis on which the physical arguments are formulated.
\par
Absent the explicit identification of a consequential error in the
derivation, Eqs. (\ref{EQt2.11}) and (\ref{EQt2.13}) can be treated as
proven.
There are no provable errors in the derivation and we should very much
like to identify the source and remedy of the inconsistencies.
This process has begun and we can report success with the Fresnel
relations \cite{BIFresnel} and special relativity in a
dielectric \cite{BIRel}.
\par
1) 
In continuum dynamics, boundary conditions are derived by the
simultaneous application of conservation of energy and conservation of
linear momentum at the boundary.
This procedure fails to work for continuum electrodynamics because the
electromagnetic energy and momentum are both quadratic in the fields.
For the electromagnetic boundary conditions and the Fresnel relations,
the established textbook derivation procedure is based on the application
of Stokes's theorem and the divergence theorem to the Maxwell--Minkowski
equations \cite{BIJackson,BIMar,BIGriffiths,BIZangwill}.
The textbook derivation procedure is specific to the Maxwell--Minkowski
formulation of continuum electrodynamics, Eqs.~(\ref{EQt2.02}), and isn't
based on conservation laws: We need a general procedure.
In Ref. \cite{BIFresnel}, the dielectric boundary conditions and Fresnel
relations were derived from the wave equation and conservation of energy.
\par
The boundary conditions and Fresnel relations that are derived from the
general procedure are the same as those derived from the
Maxwell--Minkowski equations and are equally consistent with experiments.
\par
2)
The field equations, Eqs.~(\ref{EQt2.13}) have the same form as the
Maxwell equations in the vacuum with $x^0\rightarrow \bar x^0$.
Letting $c\rightarrow c/n$, the material Lorentz factor for a linear medium
\begin{equation}
\gamma_d=\frac{1}{\sqrt{1-n^2v^2/c^2}}
\label{EQt3.01}
\end{equation}
is inconsistent with special relativity.
If Eqs.~(\ref{EQt2.13}) are proven wrong in this manner, then
Eqs.~(\ref{EQt2.02}) are proven wrong and we have a significant do-over.
\par
In adapting Einstein's special theory of relativity to a dielectric medium,
Laue \cite{BILaue,BILaue2} applied the relativistic velocity sum rule to a
dielectric material moving uniformly in a Laboratory Frame of Reference.
In Einstein--Laue dielectric special relativity, the speed of light in the
dielectric depends on the velocity of the dielectric relative to the
Laboratory Frame and the Lorentz factor is unchanged from Einstein's
special relativity.
In contrast, Ref. \cite{BIRel} derives the Rosen \cite{BIRosen} theory of
dielectric special relativity for inertial reference frames translating at
constant speed in an arbitrarily large region of space in which the speed
of light is $c/n$ in the local rest frame.
In this case, the speed of light at the location of the observer in the
dielectric is independent of the motion of the dielectric in accordance
with the principle of relativity but the material Lorentz factor,
Eq.~(\ref{EQt3.01}), is material dependent confirming the phenomenological
factor shown in Eq.~(\ref{EQt3.01}) and in Ref.~\cite{BIRosen}.
\par
The Einstein--Laue theory of dielectric special relativity is
experimentally proven by the Fizeau \cite{BIFizeau} water tube experiment.
An observer in the vacuum, such as Fizeau \cite{BIFizeau}, will measure
the speed of light in a dielectric to be dependent on the velocity of
the dielectric relative to the Laboratory Frame of Reference,
We can never place a matter-based observer, no matter how small, in a
continuous dielectric because the model dielectric is continuous at all
length scales and will always be displaced.
Ref.~\cite{BIRel} shows how to relate measurements in the vacuum to
events in the dielectric so that the Rosen theory of dielectric
special relativity is also experimentally proven.
\par
\section{Conclusion}
\par
The essential point of this communication is that the variant
formulation of macroscopic field equations, Eqs.~(\ref{EQt2.11}),
is a rigorously derived and valid identity of the macroscopic
Maxwell--Minkowski equations for a simple magneto-dielectric medium.
Consequently, any physical principles that are inconsistent with 
Eqs.~(\ref{EQt2.11}) do not prove an error in the derivation of
Eqs.~(\ref{EQt2.11}) but, instead, indicate either
{\it i)} an error in the application of the fundamental physical
principles, which are intrinsic to the vacuum, to a linear
magneto-dielectric medium,
{\it ii)} an inconsistency between the macroscopic Maxwell field
equations and the fundamental physical principles, or
{\it iii)} both.
Having proven Eqs.~(\ref{EQt2.11}) to be identities of the macroscopic
Maxwell--Minkowski equations for a simple linear medium, we can use 
these results in future work without further argument about their
validity.
\par

\end{document}